\documentclass{aa} 
\usepackage{graphicx,amssymb,psfig,natbib}

\newcommand{\AaA}{A\&A}

\newcommand{\ApJS}{Astrophysical Journal Supplement Series}
\newcommand{\ApJ}{ApJ}

\newcommand{\RvMP}{Reviews of Modern Physics}

\begin{document}

\title{Photoelectric effect on dust grains across the L1721 cloud in
        the $\rho$ Ophiuchi molecular complex
        \thanks{Based on observations with ISO, an ESA project
        with instruments funded by ESA Member States (especially the PI
        countries: France, Germany, the Netherlands and the United
Kingdom)
        and with the participation of ISAS and NASA. }}
   
\author{E. Habart\inst{1}, L. Verstraete\inst{1}, F. Boulanger\inst{1}, G. Pineau des For\^{e}ts\inst{1}, F. Le Peintre\inst{1} \and 
           J.P. Bernard\inst{1}} 
 
\offprints{E. Habart (emilie.habart@ias.u-psud.fr)} 
 
\institute{Institut d'Astrophysique Spatiale, Orsay, France} 
 
\date{Received January 12 / accepted March 08, 2001} 

\abstract{
We present ISO-LWS measurements of the main gas cooling lines, $[{\rm
C}^+]~158\mu$m and $[{\rm
O}^0]~63\mu$m towards a moderate opacity molecular cloud ($Av \sim 3$),
L1721, illuminated by
the B2 star $\nu$ Sco ($\chi$ = 5-10). These data are
combined with an extinction map and IRAS dust emission images to test
our understanding of gas heating and cooling
in photo-dissociation regions (PDRs). This nearby PDR is spatially
resolved in the IRAS images; variations in
the IRAS colors across the cloud indicate an enhanced abundance of small
dust grains within the  PDR.
A spatial correlation between the gas cooling lines and the infrared emission
from small dust grains illustrates the dominant role of small dust grains in 
the gas heating
through the photoelectric effect.
The photoelectric efficiency, determined from the observations by ratioing
the power radiated by gas and
small dust grains, is in the range 2 to 3 \% in close agreement with recent 
theoretical estimates \cite[]{bakes94,weingartner01}.
The brightness profiles across the PDR in the $[{\rm C}^+]~158\mu$m and
$[{\rm O}^0]~63\mu$m
lines are compared with model calculations where the density profile is
constrained by the extinction data
and where the gas chemical and thermal balances are solved at each position. We
show that abundance variations of small dust grains across the PDR must be
considered to account for the LWS observations.
}

\authorrunning{Habart et al.}
\titlerunning{Photoelectric efficiency across L1721}

\maketitle

\section{Introduction}

The bulk of interstellar matter is found in regions of low to moderate
opacity to UV and visible light where stellar
radiation plays a dominant role in determining the chemical and thermal
state of the gas. These
photon-dominated or photo-dissociation regions have been the subject
of many observations and theoretical studies over the past twenty years
\cite[for a review see ]{hollenbach99}.
In the classical theoretical paradigm, a PDR is characterized by its
proton density, $n_H$, and a scaling factor, $\chi$,
which normalises the incident radiation field to the Solar Neighbourhood
radiation field in the far-ultraviolet.
Comparison between observations and model calculations have concentrated
on bright objects with high $\chi > 10^3$ and
gas density $n_H > 10^3$ cm$^{-3}$ which are the easiest to observe. The gain in
sensitivity provided by the Infrared Space Observatory
(ISO) for gas line observations now permits to extend these studies to
less excited PDRs with lower $n_H$ and $\chi$ which are the most
widespread in the interstellar medium (ISM).
Models show that in such PDRs the gas heating is dominated
by the photoelectric effect on dust grains
\cite[]{hollenbach91,lebourlot93a,kemper99}.
Photoelectric heating has been the subject
of recent theoretical investigations which provide an estimate of the
photoelectric yield as a function of grain size
where the small grains (radius $\le$100 \AA)
dominate the overall photoelectric heating \cite[]{bakes94,weingartner01}.
It is thus in low excitation PDRs including the diffuse interstellar medium
that one can most directly validate this theoretical understanding of a
key physical process for the ISM,
independently of other heating processes. Recently, \cite{wolfire95} have
used the work of Bakes and Tielens (1994, hereafter BT)
to compute the equilibrium thermal states of the diffuse atomic medium.
For the Solar Neighbourhood, their calculations give
a net estimate of the  $[{\rm C}^+]~158\mu$m line emission per hydrogen
atom  in agreement with
the value derived from  the comparison of the FIRAS and H~I data at high
Galactic latitude \cite[]{boulanger96}. However, this validation of the BT work
depends on the relative gas mass in the cold
and warm phases of the ISM which is poorly constrained observationally.

In this paper, we investigate the photoelectric heating of the ISM
through the study of a specific, low-excitation PDR. The comparison of the power
radiated by gas and dust allows us to correlate the gas and
dust emission and determine the photoelectric efficiency.

To carry out this study, we selected the Lynds Dark Nebula 1721
\cite[L1721, ]{lynds62},
a nearby molecular cloud in the $\rho$ Ophiuchi region. This cloud 
as seen in the IRAS data has a roughly spherical geometry and
a  moderate opacity to stellar light ($Av = 3$ at cloud center).
It is heated on one side by the B2 star $\nu$ Sco and by a more
isotropic interstellar radiation field created by the Upper Scorpius 
association.
We present in this paper observations of the $[{\rm C}^+]~158\mu$m and
$[{\rm O}^0]~63\mu$m lines
obtained with the Long Wavelength Spectrometer
\cite[LWS, ]{clegg96a} on board the Infrared Space Observatory \cite[ISO,
]{kessler96}. These observations
are combined with IRAS images and an extinction map (Sects. 2-3).
Thanks to its proximity to the Sun and the moderate gas density, the
L1721 PDR is
spatially resolved by both the gas and dust observations.
The data is interpreted within the framework
of a model of the gas emission (Sect. 3).
The data allows us to determine the photoelectric efficiency and to
illustrate the dominant role of small dust grains
in the photoelectic heating (Sect. 4). The paper conclusions are
summarised in Sect. 5.

\begin{figure}[h!]
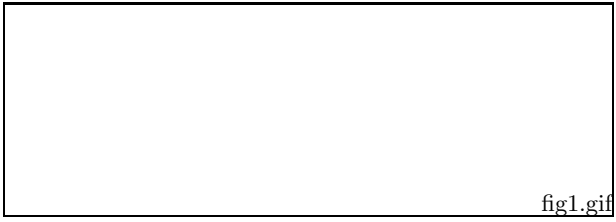


\begin{center}
\leavevmode
\framebox(230,80)[br]{fig1.gif}
\end{center}

\caption{\em L1721 as seen in the IRAS data. Top: map of the sum of IRAS bands (12+25+60+100 $\mu$m) 
with the beams of LWS observations projected on sky. Bottom: map of the 12/100 $\mu$m IRAS band ratio. In this latter map, 
the limb brightening or halo of the cloud is clearly visible. The circles show the LWS beam in the cut taken across the cloud (20 positions) and the position of $\nu$ Sco is marked by a star.}
\label{carte}
\end{figure}

\section{Observations}

The L1721 cloud is centered at $\alpha$ = 16h 14mn 30.41s and $\beta$ = -18$^{\circ}$ 58' 7.9'' (Epoch 2000), at the north 
of the $\rho$ Ophiuchi complex. 
The B2 star $\nu$ Sco is located at a distance of $\sim$134 pc.
The projected distance between the center of the cloud and the star is approximately 1.7 pc. 
We discuss in detail the incident radiation field of L1721 in Sect. 3.2.

On the map of the 12/100 $\mu$m IRAS band ratio (see Fig. \ref{carte}), the limb brightening of the cloud
 is clearly visible. Former studies have shown that ratios of the different IRAS bands mostly trace dust abundance variations \cite[]{boulanger90,bernard93}. We discuss the abundance variations of small grains across L1721 in Sect. 3.5.

\subsection{Gas cooling lines}

\begin{figure}[h!]
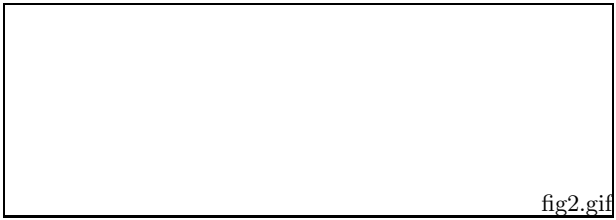


\begin{center}
\leavevmode
\framebox(230,80)[br]{fig2.gif}
\end{center}

\caption{\em Intensities of the C$^+$ and O$^0$ lines with $3\sigma$ error bars as measured by LWS at each position across L1721. The background contribution has been subtracted (see text). The $\nu$ Sco star lies at the right end of the cut.}
\label{profils}
\end{figure}

The $[{\rm O}^0]~63\mu$m and $[{\rm C}^+]~158\mu$m lines have been observed by
ISO with
the LWS (TDT = 45401125, LWS02 grating mode, Clegg et al., 1996).
At each sky position, the C$^+$ and O$^0$
lines have been observed 8 and 30 times respectively with 5 seconds-individual scans
(the integration per grating point was 0.4 seconds). 
The resolution (FWHM) is $\Delta\lambda$ = 0.5 and 0.29 $\mu$m for the C$^+$ and O$^0$ lines respectively and each resolution
element was sampled twice. We use the LWS pipeline
version 7 for standard data reduction. To remove glitches (cosmic ray impacts on the detectors), we co-add all scans to obtain a first estimate of the
mean line spectrum. Subtracting this mean spectrum from each individual scan, we obtain a (noise + glitch) - spectrum in 
which glitches are identified as 3$\sigma$- and 5$\sigma$-outliers for C$^+$ and O$^0$ respectively. 
After rejection of glitches, all scans are again co-added
to produce the final line spectrum.
Line fluxes and error bars are obtained by fitting a Gaussian and a linear baseline to the profiles. The Fig. \ref{profils} shows the line intensity profiles.

To measure the line emission from L1721, we need to subtract the contribution
of the diffuse ISM along the line of sight. 
We express the background contribution to the line emission as:
\begin{equation}
I_b = Y \times \nu I_{\nu}
\end{equation}
where $Y$ is the line-to-continuum ratio and $\nu I_{\nu}$ the continuum brightness at the line wavelength.
In the case of the C$^+$ line, we take the line-to-continuum ratio measured in the Galaxy at high latitude,
 $Y_{\rm C^+}$ = 0.009$\pm$0.0015 from \cite{boulanger96}. 
This value agrees well with the Galactic plane mean value of 0.0082 reported by \cite{wright91b} based 
on the FIRAS (Far-Infrared Absolute Spectrophotometer) survey of the far-infrared emission from our Galaxy.
\cite{wright91b} observed, as expected theoretically, that the intensity of the C$^+$ line scales linearly with the dust continuum
emission. The variations of the C$^+$ line-to-continuum ratio, due to differences in the dust properties,
in the color of the ambient interstellar radiation field or the saturation of the C$^+$ cooling rate 
when the level populations are in equilibrium with dense warm gas, are within 25\% in the plane of Galaxy.
In the following we consider that $Y_{\rm C^+}$ = 9$\pm$2.25 10$^{-3}$
in the proximity of the L1721 cloud.
For the O$^0$ line, the combination of the observations of \cite{caux97} in the diffuse ISM and the IRAS band at 60 $\mu$m 
provide a large range of $Y_{\rm O^0}$ values from 5 10$^{-4}$ to 5 10$^{-3}$.
To derive the O$^0$ background around the L1721 cloud we take the median of these values $Y_{\rm O^0}$ = 1.7 10$^{-3}$.
Note however that the large uncertainty on $Y_{\rm O^0}$ will not affect our estimate of the O$^0$ line emission in the L1721 cloud
because the contribution of the diffuse ISM to the O$^0$ emission along the line of sight 
is much lower than the line intensity in the cloud.    
For the continuum brightness, 
we use the DIRBE (Diffuse Infrared Background Experiment) data corrected for zodiacal light.
The continuum values are found from the average of DIRBE points in the diffuse ISM around the L1721 cloud. 
The variations of the DIRBE values are typically 
within 20\% in the surrounding of the cloud.
We find: $\nu I_{\nu}$(158 $\mu$m) = 9.2$\pm$1.9 10$^{-4}$ ergs/s/cm$^{2}$/sr and $\nu I_{\nu}$(60 $\mu$m) = 3.4$\pm$0.7 10$^{-4}$ ergs/s/cm$^{2}$/sr.
The corresponding background emission values which have been subtracted from our LWS cut are 8.3$\pm$2.7 10$^{-6}$ and 
5.8$\pm$1.2 10$^{-7}$ ergs/s/cm$^{2}$/sr for C$^+$ and O$^0$ respectively.

\begin{figure}[h!]
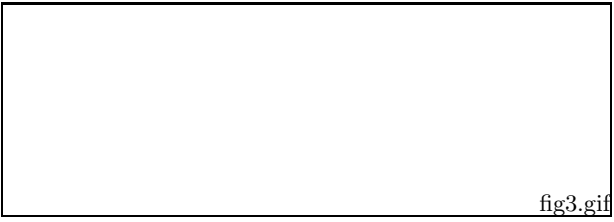


\begin{center}
\leavevmode
\framebox(230,80)[br]{fig3.gif}
\end{center}

\caption{\em Surface brightness in the 4 IRAS bands as a function of the observed LWS positions.
The background emission has been subtracted (see Sect. 2.2).}
\label{iras}
\end{figure}

\subsection{Dust infrared emission}
Fig. \ref{iras} shows the background subtracted surface brightness in the four IRAS bands. 
For each IRAS band, the background emission has been estimated along cuts extending the LWS cut at both ends of the cloud. The background emission
is taken to be the average of the minimum values of these extended cuts:
$I^b_{12}=1.2\pm0.2, I^b_{25}=1.8\pm0.3, I^b_{60}=7\pm1.6$ and $I^b_{100}=35\pm3.9$ MJy/sr.
\\
The IRAS and COBE data are well explained with dust models comprising three components \cite[]{desert90,dwek97}. We adopt
the D\'esert et al. model: the three dust components are thus by order of increasing sizes (in terms of the grain radius, $a$):
\begin{enumerate}
\item large aromatic molecules or PAHs (Polycyclic Aromatic Hydrocarbons) which contribute most of the 12 $\mu m$ band ($a$ = 4 to 12 \AA)
\item the  Very Small Grains (VSGs) which dominate the 25 $\mu m$ and 60 $\mu m$  bands ($a$ = 12 to 150 \AA) 
\item the Big Grains (BGs) which make most of the 100 $\mu m$ band ($a$ = 1.5 10$^{-2}$ to 0.11 $\mu m$).
\end{enumerate}

In the following, we will often refer to PAHs and VSGs as the {\em small dust grain} populations. 
To derive the photoelectric efficiency for each grain component, we need to decompose the total infrared (IR) emission into 
individual dust contributions at each position in the cloud.

In the case of PAHs, we use the following relationship based on ISOCAM and DIRBE data \cite[]{boulanger96a,bernard94}:\\  
\\
$I_{PAH}$ = 1.5$\times$$\nu I_{\nu}$(12 $\mu$m) erg cm$^{-2}$ s$^{-1}$ sr$^{-1}$.\\
\\
The emission of big grains is modelled with a modified blackbody of temperature $T_{BG}$ and an emissivity law 
proportional to $\nu^{2}$. The value of $T_{BG}$ is estimated using the flux measured by all ten LWS detectors covering the range
 $\lambda$ = 42.5-176 $\mu$m. At each position, we fit the ten spectral points with the modified blackbody: a constant temperature of 
$T_{BG}$ = 20 K at all positions is found to provide a satisfactory fit.
Finally, this modified blackbody is scaled to match the 100 $\mu$m IRAS band.

For the VSGs emission, we first obtain two spectral points at 25 and 60 $\mu$m from:\\
\\
$I_{25,60}(VSG) = I_{25,60}(IRAS)-I_{25,60}(BG)$\\
\\
At each cloud position, we fit these two points with a modified blackbody of constant temperature $T_{VSG}$ = 70 K and an 
emissivity law proportional to $\nu$.

\begin{figure}[h!]
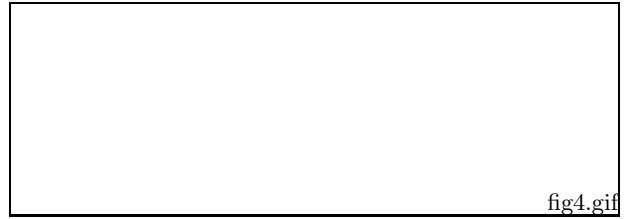


\begin{center}
\leavevmode
\framebox(230,80)[br]{fig4.gif}
\end{center}

\caption{\em Integrated flux of each dust component as a function of the position along the LWS cut. $\nu$ Sco lies at the right end of the cut.}
\label{3comp}
\end{figure}

The bolometric intensity of each dust component is shown in Fig. \ref{3comp}:
in fact, the VSGs and BGs emission profiles are quite similar to the 25+60 $\mu m$ and
100 $\mu m$ IRAS band profiles respectively.
We note that BGs and VSGs temperatures are expected to decrease as one penetrates into
the cloud. The constant temperature we find may be due to the spherical
geometry of the cloud where warm and cold dust 
are mixed along a given line of sight.

\section{Modelling the gas line cooling}

To study the photoelectric effect on dust grains across this nearby PDR,
we will compare the total gas cooling lines to the dust 
IR emission (see Sect. 4). In the case of low excitation PDRs, cooling by the pure rotational lines of H$_2$ may be important. 
As this latter has not been observed, we resort to an updated version of the PDR model described in 
Le Bourlot et al. (1993) to estimate the cooling by H$_2$. 
Moreover the comparison between the brightness profiles observed across the L1721
PDR in the $[{\rm C}^+]~158\mu$m and $[{\rm O}^0]~63\mu$m lines with model calculations provides a test for our understanding of gas heating/cooling in PDRs. The key ingredients of this model are the density profile across the PDR and the incident radiation field at the cloud surface. In the following, we describe how these quantities are 
constrained in order to produce a realistic model of the L1721 PDR.  

\subsection{The density distribution}

\begin{figure}[h!]
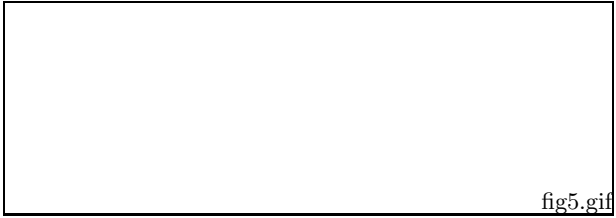


\begin{center}
\leavevmode

\framebox(230,80)[br]{fig5.gif}
\end{center}

\caption{\em Comparison of the column density observed (solid line) across L1721 and
the analytic function fitted to it (dotted line) as a function of radial distance, $r$, and the visual extinction integrated from the cloud border, $A_v$. The peak observed at the right end of the cut is due to the plate saturation by the $\nu$ Sco star. The LWS positions are marked by squares.
}
\label{extinc_iras_1}
\end{figure}

To determine the density profile of L1721 we use a visual extinction
map derived from star counts \cite[]{cambresy99}. 
Fig. \ref{extinc_iras_1} displays the column density, $N_H = Av / (0.54 \times 10^{-21})$ cm$^{-2}$ observed across L1721.
The column density profile is well represented by a gas density distribution of the form (see Fig. \ref{extinc_iras_1}):

\begin{equation}
n_H(r) = \left\{ \begin{array}{r@{\quad:\quad}l}
n^{o}_{H}\times(r/r_1)^{-1.9} & r_{1} < r \le r_{m} \\ n^{o}_{H}\hfill & r \le r_{1} 
\end{array} \right.
\end{equation}

\noindent with $r$ the radial distance to the cloud center, n$^{o}_{H}$ = 2915 cm$^{-3}$ the constant density in the $0 \le r \le r_1$ region
with $r_1$ = 0.2 pc and $r_m$ = 2.2 pc.
In this derivation, we assume that $R_v$ is constant and equal to 3.1 across the L1721 cloud. A higher value of $R_v$, 5.5, that is commonly observed in opaque cores, would multiply our extinction values by a small factor, about 1.12.

\subsection{The radiation field}

\begin{figure}[h!]
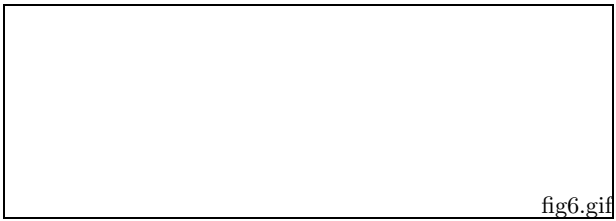


\begin{center}
\leavevmode
\framebox(230,80)[br]{fig6.gif}
\end{center}
\caption{\em The 100 $\mu$m IRAS surface brightness (a) opposite to $\nu$ Sco and (b) facing $\nu$ Sco as a function of the column density observed along each line of sight.}
\label{extinc_iras_2}
\end{figure}

The ratio $I_{100 \mu m}/N_H$ provides a good measure of the radiation field strength \cite[see Fig. 9 in ]{bernard92a}.
In Fig. \ref{extinc_iras_2}, we show the relationship between the IRAS 100 $\mu$m surface brightness and the observed column density. 
In the diffuse part of the cloud ($A_v \le 0.8$), we observe a good linear correlation
between the IRAS 100 $\mu$m surface brightness and the column density. 
The slope $I_{100 \mu m}/N_H$ of this correlation provides an estimate of the radiation field intensity.
For a set of clouds in Chamaleon with an opacity comparable to L1721 \cite{boulanger98a} found a $I_{100 \mu m}/N_H$ 
ratio of 7.5$\pm$2.5 MJy/sr/mag which corresponds to 4$\pm$1.3 MJy/sr/(10$^{21}$ cm$^{-2}$). This study
provides the reference value for the mean Solar Neighbourhood ISRF (Interstellar Standard Radiation Field).
In L1721 we find a slope of 15 $\pm 4$ for the cloud side opposite to $\nu$ Sco and 23 $\pm 4$ for the cloud side
 facing $\nu$ Sco (see Fig. \ref{extinc_iras_2}) which corresponds to 4 and 6 times the ISRF respectively.
We emphasize here that the IRAS 100 $\mu$m
emission profile is relatively insensitive to abundance variations of the small grains (PAHs and VSGs) as shown in \cite{bernard92a}.

\begin{figure}[h]
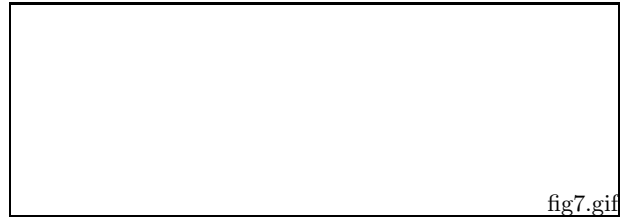

\begin{center}
\leavevmode
\framebox(230,80)[br]{fig7.gif}
\end{center}

\caption{\em Geometry used for modelling L1721. The projected distance of $\nu$ Sco to the cloud center is 1.7 pc.}
\label{dessin1} 
\end{figure}

The radiation field heating L1721 is the sum of two contributions.
The first contribution is from the $\nu$ Sco star, which has a luminosity of 6300 $L_{\odot}$ and an effective temperature of 20,000 K. The contribution of $\nu$ Sco depends 
on the position of the star along the line of sight: we reference this position with an angle $\theta$ between the sky plane
and the line connecting the cloud center to $\nu$ Sco (see Fig. \ref{dessin1}).
Next, the ISRF at this position is likely to be enhanced relative to the
mean Solar Neighbourhood ISRF because L1721 is heated by the B and A stars cluster of Upper Scorpius.
The Upper Scorpius complex contains a large proportion of stars with effective temperature of less than 20,000 K \cite[]{degeus89}. We use the \cite{mathis83} radiation field
to represent the mean Upper Scorpius radiation field.
This last radiation field is assumed to be isotropic and in the following we call this contribution the {\em isotropic radiation field}.
The radiation field exciting L1721 thus only depends on the distance 
of $\nu$ Sco to the cloud center and on the strength of the 
Upper Scorpius contribution.
The slope of the 100 $\mu$m-$N_H$ correlation measures the power absorbed by Big Grains
per Hydrogen atom. This power depends on the intensity but also on the spectral distribution
of the radiation field. In practice, for a given radiation field intensity at 1000 \AA, the 
emission of BGs per H atom is 2.5 times larger for the \cite{mathis83} radiation field than for a 20,000 K
blackbody. We have taken into account this dependence to translate the 100 $\mu$m/$N_H$
slopes into $\chi$-values which represent the strength of a radiation field at 
$\lambda$ = 1000 \AA \hspace{0.2cm} in units of 1.6 $10^{-3}$ ergs/cm$^2$/sr \cite[]{habing68}.
The radiation field intensity is described in this way because the far-UV photons are more efficient
to heat the gas.
The strength of the $\nu$ Sco radiation field is represented by $\chi_{sco}$
and the strength of the {\em isotropic radiation field} by the factor $\chi_I$. 
We have considered two cases:

\begin{enumerate}
\item $\chi_I = 1$ for the isotropic radiation field and $\chi_{sco} = 12.5$. 
In that case most of the radiation field strength on both cloud sides is due to $\nu$ Sco.
\item The Upper Scorpius isotropic radiation field on L1721 is the dominant radiation field on the cloud side opposite to $\nu$ Sco and then 
$\chi_I = 4$ and $\chi_{sco} = 5$.
\end{enumerate}

A detailed modelling of the dust IR emission in L1721, based on the model of \cite{bernard92a},
is currently underway in our group to discriminate between these two cases.
We expect that the case 2 radiation field will propagate more efficiently than the UV light 
towards the cloud center because of the strong visible component 
of the radiation field. Also, note that the effects of clumpiness can affect the penetration
of star light inside the cloud. 
For the line modelling, we select the case 2 to describe the exciting radiation field of L1721. 
In Sect. 3.4, we will discuss the sensitivity of the gas cooling lines emission to
the incident radiation field.   
In the case 2, the contribution of $\nu$ Sco amounts to a mean value of $\chi_{sco}$ = 5 for the cloud side facing $\nu$ Sco, which corresponds to $\theta = 70^{\circ}$ for the position of the star. As the far-infrared extinction is small,
a very similar 100 $\mu m$ emission band profile is expected for $\theta = -70^{\circ}$.
We will see in Sect. 3.4 that $\theta = +70^{\circ}$ is favoured.

\subsection{PDR model}

To interpret the gas line emission we make use of the model described in \cite{lebourlot93a}.
In this model a PDR is represented by a
semi-infinite plane-parallel slab with an isotropic radiation field incident on the interface.
The inputs parameters are {\rm (i)} $\chi$, the scaling factor for the 
radiation field, and {\rm (ii)} the density profile. 
With these inputs the model solves the chemical and thermal balances starting from the slab edge.
The scattering of the radiation by dust grains is treated with the formalism of \cite{flannery80} in the plane-parallel case and assumes a mostly
forward scattering ($g=<cos \alpha>$=0.6, where $\alpha$ is the scattering angle). For extinction properties, we use the analytical fit of \cite{fitzpatrick88} to the Galactic average extinction curve.
The radiative transfer in the absorption lines of H$_2$ and CO is treated in detail and the individual line
profiles are treated with the prescription of \cite{Federman79}.
We use recent gas phase elemental abundances measured in the diffuse interstellar medium: He/H = 0.1, C/H = 1.4 10$^{-4}$ 
\cite[]{cardelli96}, and O/H = 3.19 10$^{-4}$ \cite[]{meyer98}. \\
For the photoelectric effect on small dust grains, we adopt the formalism of BT. 
In particular, our model takes into account the actual spectral distribution of the radiation field to compute the heating rate
(Eqs. 1 and 14 of BT).
The exponent of the power law size distribution of small grains is -3.5 while the lower and upper limits of the grain radius are 
4 and 100 \AA \hspace{0.2cm} respectively. Larger grains are not included in the computation of the photoelectric 
heating rate. They have a low photoelectric efficiency due to the large number
of collisions the electron must undergo before escaping the grain \cite[]{verstraete90}. 
The computation of the photoelectric heating assumes that the small grains are graphitic
and spherical. Grains with less than 50 C atoms are likely to be planar: these species (PAHs) are less charged and consequently have a higher photoelectric efficiency (BT). However, for a given radiation field, the gain does not exceed 25\% \cite[]{wolfire95}:
such variations fall within our error bars on the gas cooling lines. 

\subsection{Spatial distribution of line emission}
\nopagebreak

\begin{figure}[h!]
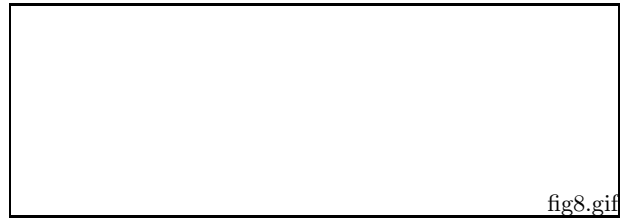


\begin{center}
\leavevmode
\framebox(230,80)[br]{fig8.gif}
\end{center}

\caption{\em Variation of the gas temperature, the gas density and the fractional ionization $x = n_e/n_H$ for a plane-parallel
PDR model with $\chi$ = 5 and the density profile of Sect. 3.1. We also show the H/H$_2$ and C$^+$/C$^0$/CO transitions 
for the same model. The $\nu$ Sco star lies to the left end of the figure.}

\label{param}
\end{figure}

\begin{figure}[h!]
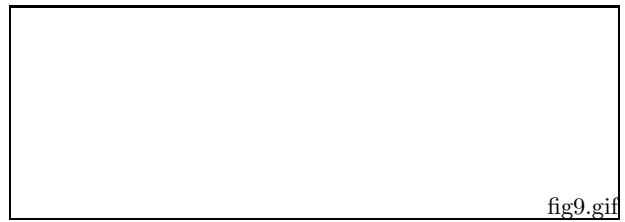


\begin{center}
\leavevmode
\framebox(230,80)[br]{fig9.gif}
\end{center}

\caption{\em Main heating and cooling rates for the PDR model ($\chi$ = 5).}
\label{bilan}
\end{figure}

We now present results of the PDR model for the parameters described in the last three sections. 
The total strength, $\chi_t = \chi_I + \chi_{sco}$, of the radiation field on the cloud surface ranges 
from 4 (opposite to $\nu$ Sco) to 11 (facing $\nu$ Sco).
Figs. \ref{param} \& \ref{bilan} describe the chemical stratification and the thermal budget of the plane-parallel model as a function of depth into the cloud. In these figures we have used $\chi = \chi_t$ = 5, which corresponds to the mean
radiation field received by a particle located on the cloud surface facing $\nu$ Sco.
We note that the photoelectric effect dominates by far the gas heat budget.

We now use the PDR model to compute profiles of line fluxes across the cloud which can be compared to the observations.
To do this, we approximate the spherical shape of the cloud by a combination of plane-parallel models.
This allows us to take into account the variation of the radiation field at the cloud surface and the fact that
each line of sight crosses diffuse, warm, interface gas as well as more deeply embedded, colder regions. 
We thus divide the cloud into angular sectors.
The first sector takes up all the cloud side opposite to $\nu$ Sco (the area noted $A$ in Fig. \ref{dessin1}).
In addition, six equal sectors are defined on the side facing $\nu$ Sco (the area noted $B$ in Fig. \ref{dessin1}): these
sectors are symmetrically distributed with respect to the line connecting $\nu$ Sco to the cloud center.
To each angular sector is associated one value of $\chi_t$. 
The three sectors facing $\nu$ Sco on one side of the 
$\nu$ Sco-cloud center line have $\chi_t$ = 11, 10  and 9 (going away from the $\nu$ Sco-cloud center line). 
The large sector $A$ opposite to $\nu$ Sco
receives  $\chi_t$ = $\chi_I$ = 4. We then run a PDR model for each sector with the density distribution of Sect. 3.1 and the
corresponding value of $\chi_t$. The model output is a local line emission rate ($J_{\nu}$) which only depends on the 
optical depth measured along a radius from the cloud edge.

\begin{figure}[h!]
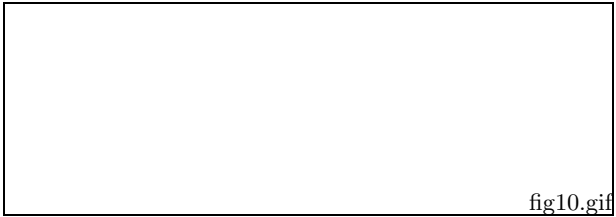


\begin{center}
\leavevmode
\framebox(230,80)[br]{fig10.gif}
\end{center}

\caption{\em Geometry used to calculate the line emission profiles intensities. Index $k$ labels the various angular sectors 
while $i$ labels the $A_v$-layers of the PDR models.}
\label{dessin2}
\end{figure}

The line intensity integrated along a given line of sight is written as (see Fig. \ref{dessin2}):   
\begin{equation}
I_{\nu} =  \sum_{k,i}^{} \frac{\Lambda_{\nu}(k,i)}{4\pi} \times l(k,i)
 \end{equation}
In this expression, $k$ labels the angular sectors and $i$ the various layers (defined by the radial $A_v$ taken from the cloud edge) crossed by the line of sight within an angular sector; $l(k,i)$ is the length in cm of the layer $i$ in the
$k$ angular sector on the line of sight. The line cooling rate in erg/s/cm$^3$ $\Lambda_{\nu}(k,i)$ is equal to $J_{\nu}(k,i) \times \beta(i)$ where $J_{\nu}(k,i)$ is the line emissivity extracted from the PDR model and $\beta(i)$ 
the escape probability from layer $i$ to the cloud edge along the line of sight. For $\beta$, we use the formalism of \cite{tielens85a} in their Appendix $B$ with a turbulent Doppler width $\delta v_d = 1$ km/s. 
We find that the $[{\rm C}^+]~158\mu$m and $[{\rm O}^0]~63\mu$m lines reach an optical depth of 1 at $A_v$=0.5 and 0.3 respectively. 
The H$_2$ rotational lines are optically thin.

\begin{figure}[h!]
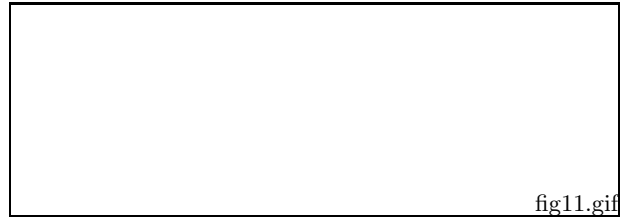


\begin{center}
\leavevmode
\framebox(230,80)[br]{fig11.gif}
\end{center}

\caption{\em Line emission profiles as observed (solid lines with squares) and predicted by the PDR model as a function of radius and of visual extinction from the cloud edge. The dotted and dashed lines correspond to a calculation with a small dust grain abundance constant throughout the cloud and $\delta v_d$ equal to 1 km/s and 3 km/s respectively. The model with small dust grain abundance variations and $\delta v_d$ = 1 km/s or 3 km/s is represented by the solid and dot-dash lines respectively.
The H$_2$ line cooling has not been observed.}
\label{emissivite}
\end{figure}

The model line intensity profiles results are compared to the data in Fig. \ref{emissivite}. We have put the star at $\theta = +70^{\circ}$ (Fig. \ref{dessin1}). Actually, for $\theta = -70^{\circ}$ the model line emission is
significantly lower due to line opacity. 
With a dust-to-gas mass ratio of the small grain populations (PAHs and VSGs) constant throughout the cloud and equal to 8.6 $10^{-4}$, the observed C$^+$ and O$^0$ line intensity profiles are not reproduced by the PDR model. Indeed, the $[{\rm C}^+]~158\mu$m and $[{\rm O}^0]~63\mu$m line intensities are underestimated by a factor of about 2 and 3 respectively. 
The uncertainty on the radiation field, mentionned in Sect. 3.2, amounts to relative error
of 40\% for C$^+$ and 70\% for O$^0$. Such error bars cannot explain the discrepancy between the observed and predicted profiles. 
Also a higher C and O abundances cannot accommodated this difference, because
the fine structure line intensities are practically linearly dependant
on the gas phase elemental abundances.  
Such discrepancies may arise from transfer effects on the emerging cooling lines and/or an underestimated heating rate. 

In the PDR model, the non-thermal velocity fields are represented by the turbulent Doppler width $\delta v_d$ which is added to the thermal width \cite[]{tielens85a}.
This parameter affects the transfer of resonant UV photons through the absorption lines of H$_2$ and CO and the opacity of the emerging IR cooling lines.
 We show the result of varying $\delta v_d$ in Fig. \ref{emissivite}: for a reasonable
range of $\delta v_d$-values, we see that decreasing the opacity of the emerging cooling lies does not account for the observed intensities, in particular $[{\rm O}^0]$.

Alternatively, the low intensity of the gas cooling lines in the model may 
tell us that the gas temperature is too low (see Fig. \ref{param}). Indeed, the predicted $[{\rm O}^0]~63\mu$m line emission is very sensitive to the gas temperature because the energy of the transition corresponds to a temperature of 228 K. Thus, another solution to match the data is to increase the heating rate.
This latter scales with the UV radiation field intensity ($\chi$), the small grains abundance and the photoelectric efficiency (BT).
The band ratios shown in Fig. \ref{abond} suggest that the PAHs and VSGs 
abundance vary across the cloud. In the next subsection we estimate these variations and show that they have an important impact on the emerging gas cooling.
We have not attempted to change the abundance of BGs, because as mentioned in the Sect. 3.3, these large grains have a low photoelectric efficiency.

\subsection{Abundance variations of the small grains}

\begin{figure}[h!]
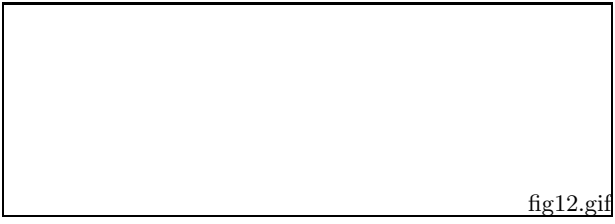


\begin{center}
\leavevmode
\framebox(230,80)[br]{fig12.gif}
\end{center}

\caption{\em Emission ratios of the different dust components as observed (solid lines with squares) and as predicted (solid lines) as a function of radius and of visual extinction from the cloud edge. }
\label{abond}
\end{figure}

In this section, we first use the dust brightness profile to quantify the abundance variations of small dust grains.
We then include these abundance variations in the PDR model and
estimate the corresponding line intensity profiles as in the last section.
Here we test the idea that the discrepancy between the observed and the predicted profiles is due to abundance variations of the small dust grains. 
Ratios of the brightness profiles of Fig. \ref{3comp} are presented in Fig. \ref{abond}.
 
\begin{figure}[h!]
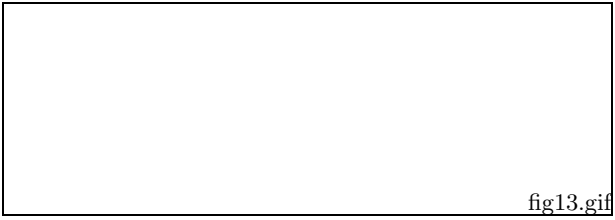


\begin{center}
\leavevmode

\framebox(230,80)[br]{fig13.gif}
\end{center}

\caption{\em Variation of the interstellar carbon abundance in PAHs (solid line) and VSGs (dotted line) of PDR models as a function of radius and of visual extinction integrated from the cloud border.}
\label{abond_profil}
\end{figure}

We have estimated the abundance variations of small grains from the ratio of PAHs and VSGs emission to that of BGs. A coherent modelling of the dust emission from a spherical cloud (e.g. Bernard et al., 1992, 1993) which includes consistent dust abundance variations is beyond the scope of the present paper.
 We estimate the model IR emission of dust grains $I^{mod}_{PAH}$, $I^{mod}_{VSG}$ and $I^{mod}_{BG}$ as follows. 
The power in the IR emission of the different dust components, $X=[PAH, VSG, BG]$, is equal to the power absorbed, {\em i.e.}:
\begin{equation}
I^{mod}_{X} = [\frac{X}{H}] \times \sum_{k}^{} \sum_{i}^{} P^X_{abs}(k,i) \times \frac{n_H(k,i)}{4\pi} \times l(k,i)
\end{equation}
where
\begin{equation}
P^X_{abs}(k,i) = \int^{13.6\;{\rm eV}}_0 \sigma_X (E) F_{k,i}(E) dE 
\end{equation}
is the power absorbed in erg s$^{-1}$ H$^{-1}$ with $\sigma_X$, the absorption cross-section per grain for PAHs, VSGs and BGs. $F_{k,i}(E)$ is the radiation field (in erg s$^{-1}$ cm$^{-2}$ eV$^{-1}$), found from the transfer prescriptions described in Sect. 3.3, and $n_H(k,i)$ the density in each cloud zone characterised by the index $k$ and $i$
(see Fig. \ref{dessin2}). For the BGs we take the cross section of \cite{desert90}, for PAHs we use \cite{verstraete92} and \cite{draine84} for VSGs.
In the ISRF, PAHs absorb 2.3 $10^{-27}$ Watt/C whereas graphitic VSGs only absorb 1.7 $10^{-27}$ Watt/C. Formula (4) can then be inverted to infer the dust abundance (C/H in PAHs and VSGs) as a function of cloud radius. The resulting abundance profiles are presented in Figs. \ref{emissivite}, \ref{abond} and \ref{abond_profil}.
The abundance of PAHs has to be multiplied by $\sim$ 5 in the $[0.25 < A_v \le 0.45]$-layer to account for the observed $I_{PAH}/I_{BG}$ ratio. 
Similarly, the VSGs abundance must be multiplied by $\sim$ 2 in the $[A_v \ge 0.75]$-layer in order to reproduce the observed $I_{VSG}/I_{BG}$ emission ratio.

\begin{table} \caption{Average power radiated per hydrogen atom by gas coolants and small dust grains towards L1721}
\begin{tabular}{lll}
\hline
 & &\\
Species \hspace{1.5cm}  Power$^a$: & Observed & Theoretical\\ 
 & &\\
\hline
\hline
 & &\\
$[{\rm C}^+]~158\mu m$ & 1.2  & 1 \\
 & &\\
$[{\rm O}^0]~63\mu m$ & 0.34 & 0.32\\
 & &\\
H$_2$ &  & 0.27 \\
 & &\\
\hline
 & &\\
Total gas cooling & 1.81$^b$ & 1.59 \\
 & &\\
$I_{PAH}+I_{VSG}$ & 78 & 54 \\
 & &\\
\hline
\hline
 & &\\
$\epsilon_{PE}$ & 0.023 & 0.029\\
  & &\\
\hline
\end{tabular}

$^a$ In 10$^{-25}$ erg s$^{-1}$ H$^{-1}$\\
$^b$ Including the H$_2$ contribution as estimated from the model
\end{table}
\normalsize

Including these abundance variations in the PDR modelling, we obtain the result
displayed in Fig. \ref{emissivite}. The corresponding line intensities are higher: the C$^+$ and O$^0$ lines emission are increased by a factor about 1.5 and 3 respectively. The discrepancy at the center of the cloud between the observed and predicted line profiles 
may result from uncertainties in the radiation field, the radiation transfer, or the photoelectric efficiency and from our rough estimate of the small dust grain abundance profiles. In the present PDR modelling we have not attempted to include changes in the UV extinction curve associated with the small dust grain abundance variations because recent observations \cite[]{boulanger94} have questioned the
contribution of small dust grains to the extinction curve adopted in the current dust models \cite[e.g. ]{desert90,dwek97}. 
In any case, exploring a range of far-UV to visible extinction ratio (from 5 to 10 at 1000 \AA) we find the cooling lines to vary by less than 50 \%. We therefore conclude, that the observed intensity profile of the main gas cooling lines can only be reproduced
if the abundance of PAHs is significantly increased. This result illustrates the dominant role of the 
smallest dust grains in the gas heating. It also shows that the model photoelectric efficiency is close to its actual value.
In Table 1, the values of the power radiated by gas coolants and small dust grains as observed and modelled are listed.

\section{Dust photoelectric efficiency}

Using our model estimate of the H$_2$ cooling, we can now estimate the photoelectric efficiency by comparing the observed dust and gas emissions. A relationship is theoretically expected between the main cooling rates and the dust emission. Indeed,
a dust grain exposed to a stellar radiation field dissipates the energy it absorbs mostly as IR emission.
A small fraction, $\epsilon$, of the energy absorbed is channelled to the gas by the photoelectric effect.
The heating efficiency of this latter process is defined as 
\begin{equation}
\epsilon = \frac{P_{PE}}{P_{abs}}
\end{equation}
where $P_{abs}$ is the total power absorbed by the dust grains and $P_{PE}$ is the power 
given to the gas through the photoelectric
effect. For a gas in thermal balance, the gas heating and cooling
 rates are equal. In the L1721 PDR, the gas heating is dominated by the photoelectric effect. Assuming thermal balance the photoelectric
heating is thus roughly equal to the total cooling rate. 
As long as the opacity of the emerging cooling lines is small, 
we can replace $P_{PE}$ by the total gas cooling line emissions $I_{\Lambda}$.
Moreover, as the power conveyed by the photoelectric
effect is much smaller than the absorbed power ($\epsilon\sim$ a few 10$^{-2}$), the latter power is well approximated by
the dust IR emission. Finally, $\epsilon$ can be observationally defined as:\\
\begin{equation}
\epsilon \simeq \frac{I_{\Lambda}}{I_{dust}}
\end{equation}
where $I_{dust}$ is the total dust IR emission.

\begin{figure}[h!]
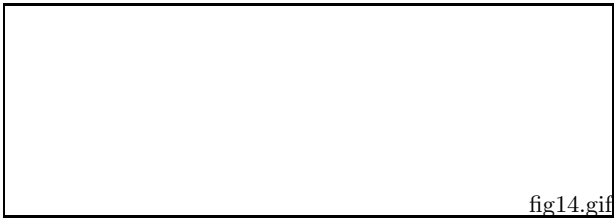


\begin{center}
\leavevmode
\framebox(230,80)[br]{fig14.gif}
\end{center}

\caption{\em Upper panel: intensities in the C$^+$ and O$^0$ lines measured by LWS, and cooling in the H$_2$
rotational lines (see Sect. 3.3.4). Lower panel: The total gas cooling, $I_{\Lambda} = I_{C^+} + I_{O^0} + I_{H_2}$  
with 3$\sigma$ error bars (solid line with crosses) and as derived from the dust IR emission: 
(i) $I^d_{\Lambda} = 0.03 \times I_{PAH}+0.01\times I_{VSG}+0.001\times I_{BG}$ (solid line) and (ii) $I^{BG}_{\Lambda} = 0.001 \times I_{PAH}+0.001\times I_{VSG}+0.02\times I_{BG}$ (dotted line).}  
\label{refcomb}
\end{figure}

\begin{figure}[h]
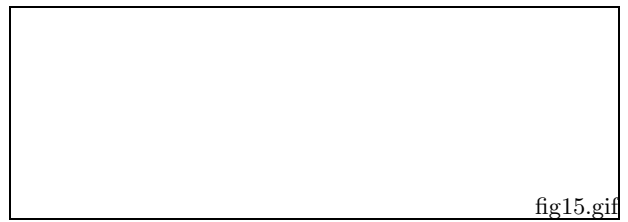


\begin{center}
\leavevmode

\framebox(230,80)[br]{fig15.gif}
\end{center}

\caption{\em Comparison between the theoretical efficiency $\epsilon_{mod}$ 
(solid line and see text) and our observational value $\epsilon_{obs} = I_{\Lambda}/(I_{PAH}+I_{VSG})$ with $I_{\Lambda} = I_{C^+}+I_{O^0}+I_{H_2}$ 
(solid line with squares) 
as a function of $r$ and $A_v$.} 

\label{efficacite}
\end{figure}

As the photoelectric heating rate is expected to be dominated by the 
small grain populations, it is interesting to determine the photoelectric efficiency for each grain population. 
Along the LWS cut, we define: 
\begin{equation}
I^d_{\Lambda} = \epsilon_{PAH}\,I_{PAH} + \epsilon_{VSG} \,I_{VSG} + \epsilon_{BG}\, I_{BG}
\end{equation}
with $\epsilon_{PAH}$, $\epsilon_{VSG}$ and $\epsilon_{BG}$ the photoelectric efficiencies of the PAHs, VSGs and BGs respectively
and $I_{PAH}$, $I_{VSG}$ and $I_{BG}$ the corresponding dust emission. 
The photoelectric efficiencies can then be found from a least-square fit of $I^d_{\Lambda}$ to $I_{\Lambda}$.
In this derivation, we assume a constant photoelectric
efficiency for each dust component across the cloud: as we show below (see Fig. \ref{efficacite}), this is a reasonable approximation.
From our best fit, we find that the small grain populations have a high photoelectric efficiency, namely
$\epsilon_{PAH}$ = 3\%, $\epsilon_{VSG}$ = 1\% and $\epsilon_{BG}$ = 0.1\% (see Fig. \ref{refcomb}).

This combination reproduces well the observed gas emission in the outer parts of the cloud,
but there is a significant deviation in the central region of the cloud.
In fact, as discussed in Sect. 3.4, the gas cooling lines become optically thick towards the cloud center
whereas the dust emission remains optically thin throughout the cloud: this is probably why $I^d_{\Lambda}$ clearly
overestimates $I_{\Lambda}$ at $r \sim$ 0.2 pc. 
We also show on Fig. \ref{refcomb} the opposite case where the BGs dominate the photoelectric heating rate:
the corresponding cooling emission $I^{BG}_{\Lambda}$ is not a good representation of $I_{\Lambda}$ along the whole LWS cut.
Our results clearly show that the dust photoelectric heating is dominated by the smaller grains,
and not by the BGs. 
However, the spatial correlation between dust and gas emission does
not allow us to tightly constrain the photoelectric efficiency of
the various grain components because of the spherical geometry where outer and inner regions 
contribute to the emergent intensity at all sky positions.

We can estimate the observational photoelectric efficiency profile of small grains across the cloud as $\epsilon_{obs} 
= I_{\Lambda}/(I_{PAH}+I_{VSG})$. This efficiency is compared in Fig. \ref{efficacite} to
the theoretical curve, $\epsilon_{mod}$ which has been derived from our PDR modelling with the dust abundance variations discussed
in Sect. 3.5. For consistency with the observations, we define $\epsilon_{mod}$ as the ratio between the emergent gas cooling lines
and the power absorbed by small grains as computed in Sect. 3.5. 
We see on Fig. \ref{efficacite} that $\epsilon_{mod}$ reproduces the main features of $\epsilon_{obs}$ across the cloud
and that the $\epsilon_{mod}$-values are within 30\% of those of $\epsilon_{obs}$.
The shape of the $\epsilon_{mod}$-profile can be understood as follows.
While penetrating into the cloud ($r \ge$0.3 pc), the UV radiation is attenuated
and the average grain charge (positive) decreases; consequently, $\epsilon_{mod}$ increases because of the weaker Coulomb barrier 
for the ejection of electrons from the grain (BT). 
Towards the center of the cloud ($r \le$0.3 pc), the photoelectric efficiency decreases.
This trend is the combination of two effects. On the one hand, the opacity of the cloud center to UV photons is larger
than for visible photons: since the photoelectric effect is produced mostly by UV photons, $\epsilon_{mod}$ diminishes
towards the cloud center. On the other hand, the $\epsilon_{mod}$-drop at $r \le$0.3 pc also reflects the behaviour
of the $n_{PAH}/n_{VSG}$ ratio (see Fig. \ref{abond}). 
Comparing $\epsilon_{mod}$ and $\epsilon_{obs}$ in detail, we see that the efficiency values at the cloud boundary are quite comparable.
Moving into the cloud, $\epsilon_{obs}$ is systematically lower than $\epsilon_{mod}$. Also, the ridge and trough of the $\epsilon_{mod}$-profile are somewhat shifted with respect to $\epsilon_{obs}$.    
These discrepancies probably result from the simplified geometry and radiation transfer we adopted for the L1721 cloud,
as well as the uncertainty in the spectral distribution of the exciting radiation field (see Sect. 3.2).
In this latter case, a detailed modelling of the dust infrared emission is warranted in
order to consistently derive the exciting radiation field and the dust abundance profiles in the L1721 cloud.

\section{Conclusion}

We study the gas thermal balance and the dust IR emission across L1721 a nearby, isolated molecular cloud heated by the $\nu$ Sco star and by the BA star association of Upper Scorpius.
The cooling of the gas is obtained with ISO-LWS measurements of the major cooling lines, $[{\rm C}^+]~158\mu m$ and $[{\rm O}^0]~63\mu m$ . The dust IR emission is traced with IRAS data. \\
We model the gas cooling lines with an updated version of the photodissociation region model of \cite{lebourlot93a}.
The input physical conditions in this modelling are the density profile and the exciting radiation field.
The density profile of the L1721 cloud is deduced from visual extinction data ($A_v$ at the center is $\sim$3).
The exciting radiation field is estimated to be $\chi$ = 5 to 10 times the radiation field of the Solar Neighbourhood
from the ratio of dust emission to gas column density.\\
The spatial correlation between the gas cooling lines and the IR emission of small grains (PAHs and VSGs of radii $\le 100$ \AA) 
confirms the theoretical expectation that the gas heat budget, in the less excited PDRs, is dominated by the photoelectric 
effect on small grains. 
The photoelectric efficiency measured from the ratio of the total gas cooling
to the dust IR emission is in the range 2 to 3\% in good agreement with current theoretical models \cite[]{bakes94,weingartner01}.\\
Moreover, we find that the gas cooling lines of the L1721 cloud cannot be explained
if the abundance of small grains is kept constant across the cloud, equal to the abundance of the diffuse ISM.
Rather, as indicated by the IRAS band ratios, our PDR model results matches the gas cooling line observations
when the abundance of PAHs is enhanced by a factor of 5
towards the cloud center.  

\acknowledgements{ We are grateful to L. Cambr\'esy for providing us the
extinction map of L1721. We also thank E. Caux and W.T. Reach for their help with the LWS data reduction.}


\begin{thebibliography}{{Weingartner} \& {Draine}(2001)}

\bibitem[{Bakes} \& {Tielens}(1994)]{bakes94}
{Bakes}, E. L.~O \& {Tielens}, A. G. G.~M.
\newblock 1994, {\em \ApJ}, 427:822.

\bibitem[{Bernard} et~al.(1992)]{bernard92a}
{Bernard}, J.~P, {Boulanger}, F, {Desert}, F.~X, \& {Puget}, J.~L.
\newblock 1992, {\em \AaA}, 263:258.

\bibitem[{Bernard} et~al.(1993)]{bernard93}
{Bernard}, J.~P, {Boulanger}, F, \& {Puget}, J.~L.
\newblock 1993, {\em \AaA}, 277:609.

\bibitem[{Bernard} et~al.(1994)]{bernard94}
{Bernard}, J.~P, {Boulanger}, F, {D\'esert}, F.~X, {Giard}, M, {Helou}, G, \&
  {Puget}, J.~L.
\newblock 1994, {\em \AaA}, 291:L5.

\bibitem[{Boulanger} et~al.(1990)]{boulanger90}
{Boulanger}, F, {Falgarone}, E, {Puget}, J.~L, \& {Helou}, G.
\newblock 1990, {\em \ApJ}, 364:136.

\bibitem[{Boulanger} et~al.(1994)]{boulanger94}
{Boulanger}, F, {Prevot}, M.~L, \& {Gry}, C.
\newblock 1994, {\em \AaA}, 284:956.

\bibitem[{Boulanger} et~al.(1996a)]{boulanger96}
{Boulanger}, F, {Abergel}, A, {Bernard}, J.~P, {Burton}, W.~B, {D\'esert},
  F.~X, {Hartmann}, D, {Lagache}, G, \& {Puget}, J.~L.
\newblock 1996a, {\em \AaA}, 312:256.

\bibitem[{Boulanger} et~al.(1996b)]{boulanger96a}
{Boulanger}, F, {Reach}, W.~T, {Abergel}, A, {Bernard}, J.~P, {Cesarsky}, C.~J,
  {Cesarsky}, D, {D\'esert}, F.~X, {Falgarone}, E, {Lequeux}, J, {Metcalfe}, L,
  {Perault}, M, {Puget}, J.~L, {Rouan}, D, {Sauvage}, M, {Tran}, D, \&
  {Vigroux}, L.
\newblock 1996b, {\em \AaA}, 315:L325.

\bibitem[{Boulanger} et~al.(1998)]{boulanger98a}
{Boulanger}, F, {Bronfman}, L, {Dame}, T.~M, \& {Thaddeus}, P.
\newblock 1998, {\em \AaA}, 332:273.

\bibitem[{Cambr\'esy}(1999)]{cambresy99}
{Cambr\'esy}, L.
\newblock 1999, {\em \AaA}, 345:965.

\bibitem[{Cardelli} et~al.(1996)]{cardelli96}
{Cardelli}, J.~A, {Meyer}, D.~M, {Jura}, M, \& {Savage}, B.~D.
\newblock 1996, {\em \ApJ}, 467:334.

\bibitem[{Caux} \& {Gry}(1997)]{caux97}
{Caux}, E \& {Gry}, C.
\newblock Iso lws spectroscopy of the diffuse interstellar medium.
\newblock In {\em The Far Infrared and Submillimetre Universe}, page~67, 1997.

\bibitem[{Clegg} et~al.(1996)]{clegg96a}
{Clegg}, P.~E, {Ade}, P.~A, {Armand}, C, {Baluteau}, J.~P, \& {Barlow}, M.~J.
\newblock et al. 1996, {\em \AaA}, 315:L38.

\bibitem[{de Geus} et~al.(1989)]{degeus89}
{de Geus}, E.~J, {DeZeeuw}, P.~T, \& {Lub}, J.
\newblock 1989, {\em \AaA}, 216:44.

\bibitem[{D\'esert} et~al.(1990)]{desert90}
{D\'esert}, F.~X, {Boulanger}, F, \& {Puget}, J.~L.
\newblock 1990, {\em \AaA}, 237:215.

\bibitem[{Draine} \& {Lee}(1984)]{draine84}
{Draine}, B.~T \& {Lee}, H.~M.
\newblock 1984, {\em \ApJ}, 285:89.

\bibitem[{Dwek} et~al.(1997)]{dwek97}
{Dwek}, E, {Arendt}, R.~G, {Fixsen}, D.~J, {Sodroski}, T.~J, {Odegard}, N,
  {Weiland}, J.~L, {Reach}, W.~T, {Hauser}, M.~G, {Kelsall}, T, {Moseley},
  S.~H, {Silverberg}, R.~F, {Shafer}, R.~A, {Ballester}, J, {Bazell}, D, \&
  {Isaacman}, R.
\newblock 1997, {\em \ApJ}, 475:565.

\bibitem[{Federman} et~al.(1979)]{Federman79}
{Federman}, S.~R, {Glassgold}, A.~E, \& {Kwan}, J.
\newblock 1979, {\em \ApJ}, 227:466.

\bibitem[{Fitzpatrick} \& {Massa}(1988)]{fitzpatrick88}
{Fitzpatrick}, E.~L \& {Massa}, D.
\newblock 1988, {\em \ApJ}, 328:734.

\bibitem[{Flannery} et~al.(1980)]{flannery80}
{Flannery}, B.~P, {Roberge}, W, \& {Rybicki}, G.~B.
\newblock 1980, {\em \ApJ}, 236:598.

\bibitem[{Habing}(1968)]{habing68}
{Habing}, H.~J.
\newblock 1968, {\em Bull.Astron.Inst.Netherlands}, 19:421.

\bibitem[{Hollenbach} \& {Tielens}(1999)]{hollenbach99}
{Hollenbach}, D.~J \& {Tielens}, A. G. G.~M.
\newblock 1999, {\em \RvMP}, 71:173.

\bibitem[{Hollenbach} et~al.(1991)]{hollenbach91}
{Hollenbach}, D.~J, {Takahashi}, T, \& {Tielens}, A. G. G.~M.
\newblock 1991, {\em \ApJ}, 377:192.

\bibitem[{Kemper} et~al.(1999)]{kemper99}
{Kemper}, C, {Spaans}, M, {Jansen}, D.~J, {Hogerheijde}, M.~R, {vansDishoeck},
  E.~F, \& {Tielens}, A. G. G.~M.
\newblock 1999, {\em \ApJ}, 515:649.

\bibitem[{Kessler} et~al.(1996)]{kessler96}
{Kessler}, M.~F, {Steinz}, J.~A, {Anderegg}, M.~E, {Clavel}, J, {Drechsel}, G,
  {Estaria}, P, {Faelker}, J, {Riedinger}, J.~R, {Robson}, A, {Taylor}, B.~G,
  \& {Ximenez De Ferran}, S.
\newblock 1996, {\em \AaA}, 315:L27.

\bibitem[{Le Bourlot} et~al.(1993)]{lebourlot93a}
{Le Bourlot}, J, {Pineau Des Forets}, G, {Roueff}, E, \& {Flower}, D.~R.
\newblock 1993, {\em \AaA}, 267:L233.

\bibitem[{Lynds}(1962)]{lynds62}
{Lynds}, B.~T.
\newblock 1962, {\em \ApJS}, 7:L1.

\bibitem[{Mathis} et~al.(1983)]{mathis83}
{Mathis}, J.~S, {Mezger}, P.~G, \& {Panagia}, N.
\newblock 1983, {\em \AaA}, 128:212.

\bibitem[{Meyer} et~al.(1998)]{meyer98}
{Meyer}, D.~M, {Jura}, M, \& {Cardelli}, J.~A.
\newblock 1998, {\em \ApJ}, 493:222.

\bibitem[{Tielens} \& {Hollenbach}(1985)]{tielens85a}
{Tielens}, A. G. G.~M \& {Hollenbach}, D.
\newblock 1985, {\em \ApJ}, 291:722.

\bibitem[{Verstraete} \& {L\'eger}(1992)]{verstraete92}
{Verstraete}, L \& {L\'eger}, A.
\newblock 1992, {\em \AaA}, 266:513.

\bibitem[{Verstraete} et~al.(1990)]{verstraete90}
{Verstraete}, L, {L\'eger}, A, {D'Hendecourt}, L, {Defourneau}, D, \& {Dutuit},
  O.
\newblock 1990, {\em \AaA}, 237:436.

\bibitem[{Weingartner} \& {Draine}(2001)]{weingartner01}
{Weingartner}, J.~C \& {Draine}, B.~T.
\newblock 2001, {\em astro-ph/9907251 submitted to \ApJ}.

\bibitem[{Wolfire} et~al.(1995)]{wolfire95}
{Wolfire}, M.~G, {Hollenbach}, D, {McKee}, C.~F, {Tielens}, A. G. G.~M, \&
  {Bakes}, E. L.~O.
\newblock 1995, {\em \ApJ}, 443:152.

\bibitem[{Wright} et~al.(1991)]{wright91b}
{Wright}, E.~L, {Mather}, J.~C, {Bennett}, C.~L, {Cheng}, E.~S, \& {Shafer},
  R.~A.
\newblock et al. 1991, {\em \ApJ}, 381:200.

\end{thebibliography}
\end{document}